# An advanced three-axis elliptical hohlraum for indirectly driven inertial confinement fusion

Longfei Jing[1], Shaoen Jiang[1,2,a)], Longyu Kuang[1,a)], Hang Li[1], Lu Zhang[1], Liling Li[1], Zhiwei Lin[1], Jianhua Zheng[1], Feng Hu[3], Yunbao Huang[4], Tianxuan Huang[1], and Yongkun Ding[1,2]

[1]Research Center of Laser Fusion, China Academy of Engineering Physics, Mianyang 621900, China

[2]Center for Applied Physics and Technology, Peking University, Beijing 100871, China

[3]School of Mathematic and Physical Science, Xuzhou Institute of Technology, Xuzhou 221111, China

[4]Mechatronics School of Guangdong University of Technology, Guangzhou 510080, China

Abstract:

The radiation symmetry and laser-plasma instabilities (LPIs) inside the conventional cylindrical hohlraum configuration are the two daunting challenges on the approach to ignition in indirectly driven inertial confinement fusion. Recently, near-vacuum cylindrical hohlraum (NVCH) [L. F. Berzak Hopkins *et al*. Phys. Rev. Lett. 114, 175001 (2015)], octahedral spherical hohlraum (SH) [K. Lan *et al*. Phys. Plasmas 21, 010704 (2014)] and novel three-axis cylindrical hohlraum (TACH) [L. Y. Kuang *et al*. Scientific Reports 6, 34636 (2016)] were proposed to mitigate these issues. While the coupling efficiency might still be a critical risk. In this paper, an advanced three-axis elliptical hohlraum (TAEH) is proposed to make a compromise among these hohlraum performance. Preliminary simulations indicate that the TAEH (with a case-to-capsule ratio, CCR=2.8) could provide excellent radiation symmetry during the thorough laser pulse of the 'high-foot' drive [J. Lindl *et al*. Phys. Plasmas 21, 020501(2014)], comparable to the ones inside the SH (CCR=5.1) and TACH (CCR=2.2). The filling time of plasma affecting the LPIs is between those of SH and TACH, and about 1.5 times of that in the ignition hohlraum Rev5-CH of NIC [S. W. Haan *et al*. Phys. Plasmas 18, 051001(2011)] and close to the one inside the NVCH (CCR=3.4). In particular, the coupling efficiency is about 22%, 29% and 17% higher than the one inside the NVCH, SH and TACH, respectively. It would be envisioned that the proposed hohlraum configuration merits consideration as an alternative route to indirect-drive ignition,

---

a) Authors to whom correspondence should be addressed. Electronic addresses: jiangshn@vip.sina.com and kuangly0402@sina.com.

complementary to the traditional cylindrical hohlraum and the proposed recently novel hohlraums.

## 1. Introduction

A high-$Z$ cavity case- namely a hohlraum, plays a significant role in indirect-drive approach to inertial confinement fusion (ICF)[1-5], which converts the energy of the incident laser beams to thermal X-rays inside the enclosing hohlraum[6]. These X-rays are repeatedly absorbed and reemitted by the hohlraum walls and deposit their energy upon the surface of a spherical fusion capsule centrally located inside the hohlraum in a nearly perfectly symmetric way. Therefore, the hohlraum configuration must be carefully designed to provide the necessary symmetry of capsule illumination[2,7]. In addition, a high enough drive on the capsule from X-rays and less laser plasma instabilities (LPIs) are also required to sharply compress the fuel capsule[2,8]. Traditionally, these hohlraums have utilized a cylindrical geometry with two laser entrance holes (LEHs) and multi-cone beams, i.e., inner cone and outer cone beams, which are adopted as the point target of National Ignition Campaign (NIC)[9-11] on the National Ignition Facility (NIF), and have achieved a great progress[10-13]. The greatest challenges on the approach to ignition utilizing the cylindrical hohlraum configuration might reside in the drive symmetry and the LPIs, such as the considerable Stimulated Raman Scatter (SRS) of the inner-cone beams[14] and cross-beam energy transfer (CBET) between the inner-cone and outer-cone beams[15] near the LEHs, which would degrade the time-dependent drive symmetry[11] and the resulting capsule implosion performance[16].

Up till to now, diverse spherical hohlraums, upon which two[17-18], three[19], four[17,20-26], six[27-30], or twelve[17] holes are drilled, with several sets of laser beams with different incident angles, have been studied on the Iskra-5, OMEGA, or NIF laser facilities. In particular, a larger and longer near-vacuum cylindrical hohlraum with lower gas fill density is also designed to improve the hohlraum performance on NIF[31-32]. However, the overlap and cross of laser beams on the LEHs plane may arouse nonlinear phenomena and complex laser-plasma interaction issues[33], such as the daunting SRS and CBET. Besides, the high-$Z$ bubbles of outer-cone beams limit inner beams propagation and low-mode symmetry control[34]. These key challenges might still remain in the rugby-like hohlraums[35-39] with two LEHs and multi-cone beams.

Therefore, developing hohlraum designs with a more symmetric and more predictable radiation environment is strongly indicated and efforts are under way to address both of these challenges[40]. Recently, a spherical hohlraum with 6LEHs[41] and a novel hohlraum with three-axis[42-43] or four-half[44] cylindrical hohlraum, all of which utilize a single-cone beams without any supplementary technology of beam phasing, are proposed to improve these issues, especially the propagation of the inner-cone beams. These hohlraum configurations have natural superiority in maintaining high drive symmetry during the entire capsule implosion process[42-45]. Notwithstanding, the laser arrangement designed for the four-half cylindrical hohlraum, which requires a greater target chamber for the ports of laser beams, might be more complicated than that for the six-end injection octahedral hohlraum. Besides, additional LEHs in spherical hohlraum and three-axis cylindrical hohlraum will increase the radiation loss through the LEHs and lead to a lower coupling efficiency from the hohlraum to capsule. According to Amendt's analysis[46], reducing the hohlraum surface area by 20% can lead to a saving of ≈100 kJ for a mega-joule class laser such as the NIF. In addition, Laborde's investigation[39] has demonstrated that an elliptical configuration could mitigate the LPIs in the two-end injection rugby-like hohlraum.

Based on the investigations stated above, an advanced Three-Axis Elliptical Hohlraum (TAEH) is proposed to balance tradeoffs among the drive symmetry, coupling efficiency, and plasma filling affecting the LPIs of the hohlraum performance for indirectly driven ICF.

The outline of this paper is as follows. Section 2 describes the hohlraum configuration. The time-dependent drive symmetry inside the configuration is analyzed in Section 3. Then Section 4 presents a comparison of the performance of assorted configurations. At last, Section 5 summarizes our discussions and conclusions.

## 2. Hohlraum configuration

As shown in Fig.1 (a) and (b) , the proposed hohlraum configuration is constructed of three orthogonal elliptical hohlraum. The 48-quads beams arrangement designed specially (see Fig.1(c)), like the hohlraum itself, has inherent octahedral symmetry with 6 quads introduced at each of the six LEHs, with the same angle of 55° to the normal of the respective LEH. Each single-cone laser beams have rotation symmetry about an axis extending from target center through the LEH center.

The laser arrangement is analogous to the one designed specially for the octahedral spherical hohlraum[47].

The shape of the laser spots focusing on the LEHs plane is round of 1.2 mm in diameter at the LEHs with diameter of 2.4 mm[41,48]. The size of the LEHs is optimized to balance tradeoffs among the laser intensity, the possibility of clipping the LEHs, and the desire to lessen the radiation loss through the LEHs[9]. The diameter of the LEHs adopted here is just for convenience to be compared with the octahedral spherical hohlraum and the three-axis cylindrical hohlraum in Section 4, where larger LEHs will be analyzed. A peak power of each laser beam is set to 10 TW artificially, so the laser intensity on the LEH plane is about $8.8 \times 10^{14}$ W/cm$^2$, which is close to the one of the outer-cone beams of the NIF[49].

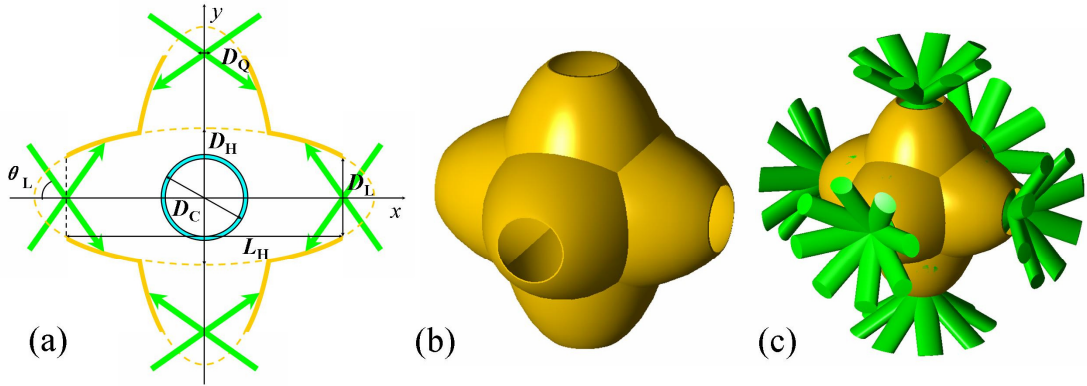

Figure 1. Schematic drawing of the Three-Axis Elliptical Hohlraum, named as TAEH. (a): two dimensional hohlraum configuration ; (b) three dimensional target fabrication; (c): layout of the target and single-cone laser beams with the incident angle of 55 degree.

## 3. Time-dependent drive symmetry

A three-dimensional view-factor code of IRAD3D[50-51] is utilized to investigate the drive symmetry. Of necessity, this code makes some simplifying assumptions and approximations about the hohlraum and the capsule. Especially, it does not include the detailed hydrodynamic evolution and radiation physics of the hohlraum and the capsule. Reference to the treatment of Ref.[42] and [48], two special sections are divided on the interior of the hohlraum, i.e., the laser spots and the wall, which are assumed as separate homogeneous background by neglecting flux difference within each section. In particular, IRAD3D can rapidly provide insight into the drive symmetry on the capsule under a variety of circumstances[52] and has been used to hohlraum shape optimization[35,53] and experimental data interpretation[54] in ICF.

For comparison, the diameter of the capsule and the LEHs are fixed at 2.2 mm and 2.4 mm[41,48] (as stated in Sec.2), respectively. The time-varying drive symmetry of the capsule is dependent on the ratio of the laser spots and the wall[55-56], namely $F_S/F_W$, which is defined by the time-dependent albedo (the ratio of the re-emitted flux to the incident flux) and the surface area of the laser spots, wall, LEHs, and the capsule[42,55]:

$$\frac{F_S}{F_W} = 1 + \frac{(1-\alpha_W)A_W + (1-\alpha_C)A_C + (1-\alpha_L)A_L}{\alpha_W A_S} \quad (1)$$

where $A_{W,C,L,S}$ are the areas of the wall, capsule, LEHs, and all the laser spots, respectively, and $\alpha_{W,C,L}$ are the albedos of the wall, capsule, and LEHs, respectively. Here $\alpha_L$ is equal to zero in fact, for all the flux passing through the LEHs will escape from the hohlraum. And we assumed the capsule albedo of $\alpha_C = 0.3$ independent of time[57] and the capsule size remained constant duration of the simulation[58]. According to the Lindl's investigation[2,59], the wall albedo $\alpha_W$ was calculated by $\alpha_W = 1 - 0.32 T_{R,heV}^{-0.7} \tau_{ns}^{-0.38}$, where $\tau$ is the time of ns and $T_{R,hev}$ is radiation temperature inside the hohlraum designed for 'high-foot' laser pulse[11] in unit of heV, as shown in Fig. 2(a). It merits mentioning that $\alpha_W$ has been set to 0.01 when $\alpha_W < 0.01$ at the initial stage. Another factor affecting the time-dependent symmetry is the locations of the laser spots, which would be altered by the motion of the hohlraum wall during a relatively long laser drives. Thereby this can compromise symmetry control[9,60]. Lindl's investigation[59] indicated that the inward wall motion was dominated by re-radiated X-rays ablation, not by laser ablation. This also means the wall can be treated as moving radially inward at the same speed throughout the hohlraum, as shown in Fig. 2(b). Although in reality the wall motion is not uniform, and the hohlraum distorts more locally under the laser spots than elsewhere; to main axial symmetry, this effect has been ignored in simulation[61]. In addition, the velocity of the wall motion is assumed to be proportional to an isothermal sound speed for Au[2], i.e., $c_{s,\mu m/ns} = 30 T_{R,heV}^{0.8} \rho_{g/cm^3}^{-0.07}$, while the slight dependence on the density is ignored here[42]. The final distance of wall shrinking was assumed to be 300 μm[61-62] and the time-varying distance $d_t$ is exhibited in Fig.2(a).

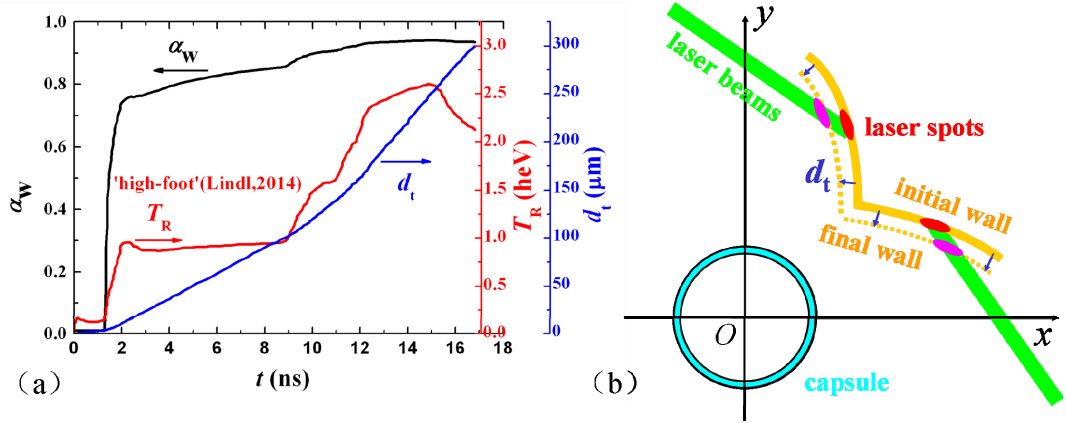

Figure 2. Schematic diagram of (a) time-dependent wall albedo, radiation temperature, distance of hohlraum wall motion and (b) spots motion due to wall motion.

The temporal evolution of drive asymmetry on the capsule inside the TAEH with different case-to-capsule ratios (CCR=2.5~3.1), is shown in Fig.3, where the criterion $\delta F$ of drive asymmetry is defined as $\delta F=0.5*|F_{max}-F_{min}|/<F>$, where the $<F>$ is the average radiation flux upon the capsule[41]. The simulation indicates that the proposed configuration almost maintains high drive symmetry at the level of less than 1% during the entire capsule implosion process[42-45], which is below the constraint of the ignition requirement to the drive asymmetry[2,59,63]. In addition, with the increase of the CCR, the asymmetry becomes less sensitive to time. While larger CCR means lower coupling efficiency. Tradeoff between the symmetry and the coupling efficiency, an initial design point with CCR=2.8 is adopted.

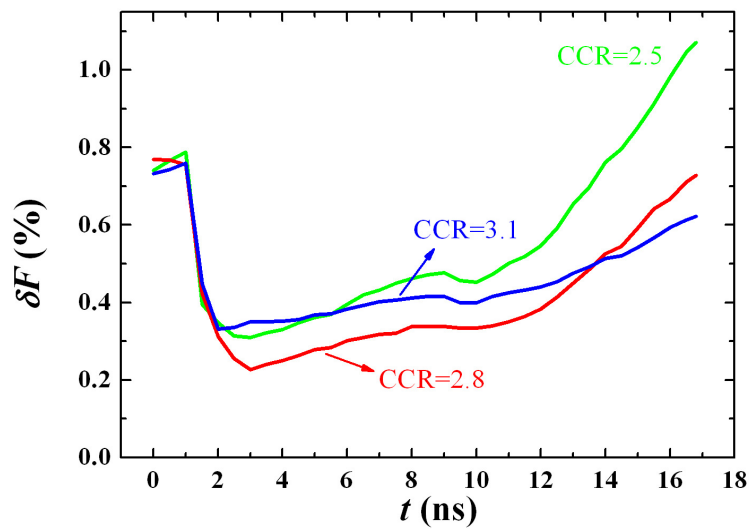

Figure 3. Temporal evolution of drive asymmetry inside the TAEH with different CCRs.

The distribution of the normalized flux incidence on the capsule at different moments inside the TAEH with CCR=2.8 is illustrated in Fig.4, which indicates a rather symmetric drive on the whole and presents intrinsic octahedral asymmetry. The relative intensity of incident flux on the capsule is varying with time. At the initial stage, the contribution of flux incidence on the capsule is dominated by the laser spots, so the zone A facing the negative source of the LEH zone displayed in Fig.1, is a little weaker than elsewhere. The motion of the laser spots due to the motion of the wall (shown in Fig.2(b)) compensates the loss from the LEHs, thereby the zone A becomes hotter and hotter. As a result, the time-integrated asymmetry could be constrained at a rather low level.

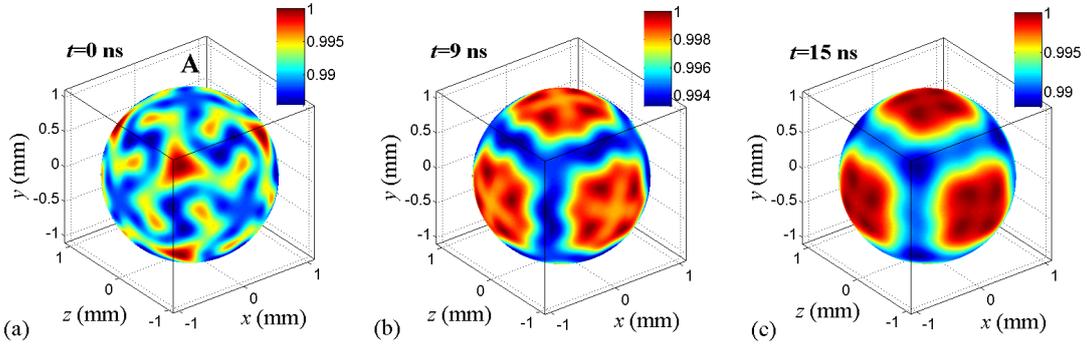

Figure 4. Distribution of normalized incident flux on the capsule at the time of (a) $t=0$ ns, (b) $t=9$ ns, and (c) $t=15$ ns, where "A" denotes the zone facing one of the six LEHs (CCR=2.8).

Fig.5 plots the dependence of major spherical harmonic components $C_{lm}$ of drive asymmetry on time. The $C_{lm}$ are defined according to Ref.[41], i.e., $C_{l0}=|f_{l0}/f_{00}|$ and $C_{lm}=2|f_{lm}/f_{00}|$ for m>0, where $f_{lm}$ are the coefficients of spherical harmonic decomposition from $F(\theta,\varphi)=\sum_{l=0}^{\infty}\sum_{m=-l}^{l}f_{lm}Y_{l}^{m}(\theta,\varphi)$, in which $F(\theta,\varphi)$ means the flux distribution upon the capsule and $Y_{l}^{m}(\theta,\varphi)$ is the spherical harmonics defined in quantum mechanics[63]. The calculations indicate that the hohlraum could provide a rather good symmetry during the whole drive period, and all the $C_{lm}$ stay below the constraint of much less than 1% .The early-time drive asymmetry on the capsule is mainly dominated by $C_{64}$ and $C_{44}$. The modes $C_{40}$ and $C_{44}$ decrease with time remarkably to reach a trough near 2.5 ns or 3.5 ns, and then increase with time gradually. In addition, these two modes dominate the asymmetry after 6 ns, and show less sensitivity to time, which accords with the simulation of Ref. [42]. The reason is that the $C_{40}$ and $C_{44}$ are dominated by the inherent hohlraum structure and laser-cone distribution of fourfold rotation symmetry, as shown in Fig.1 and Fig.4. In particular, the mode $C_{20}$ is on a noise level and could be thoroughly neglected[41], which is a significant issue needing to be mitigated in the cylindrical hohlraum[9-12]. All the other modes are constrained at a negligible level of less than 0.1%.

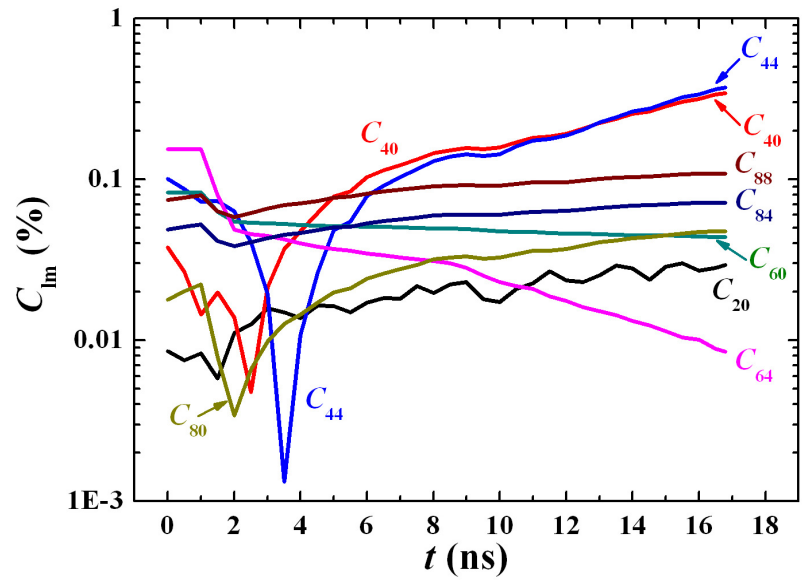

Figure 5. Variations of spherical harmonic modes of drive asymmetry as time inside the TAEH (CCR=2.8).

## 4. Performance comparison of various configurations

A slight change in the shape of a hohlraum can improve the hohlraum performance[35]. There are three factors impacting the hohlraum performance: the drive symmetry, coupling efficiency, and plasma filling affecting the laser-plasma instabilities (LPIs), which depend on the hohlraum configuration. Recently, various novel near-vacuum cylindrical hohlraum[31-32] (denoted by NVCH), spherical hohlraum[41] (denoted by SH) and three-axis cylindrical hohlraum[42-43] (denoted by TACH), as shown in Fig.6, have been proposed to improve the hohlraum performance, especially the inner beam propagation and the resulting drive symmetry. The performance of assorted hohlraum configurations is compared in this section. For comparison, the total number and power of the laser beams are assumed to be the same.

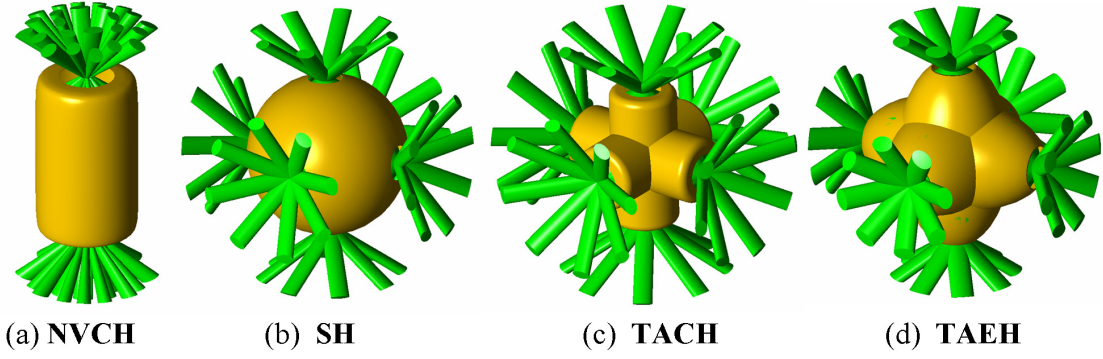

(a) NVCH    (b) SH    (c) TACH    (d) TAEH

Figure 6. Layouts of assorted hohlraum configurations, i.e., (a) near-vacuum cylindrical hohlraum (denoted by NVCH), (b) spherical hohlraum (denoted by SH), (c) three-axis cylindrical hohlraum (denoted by TACH), and (d) three-axis elliptical hohlraum (denoted by TAEH).

Aside from the drive symmetry studied above, another two criterions of coupling efficiency and filling time of plasma are adopted to characterize the hohlraum performance. A higher energy coupling will economize the input energy, increase the fusion energy gain[48], and provide a wider design margin for a larger hohlraum to gain higher symmetry[32]. Based on the simplifying approximation of spatially uniform in albedo[22] and radiation temperature[2] throughout the hohlraum, where the variations within the hohlraum are relatively small for typical laser-driven hohlraum[2], the power balance inside a hohlraum in steady-state conditions can be stated as[21,64]

$$(1-\eta_s)\eta_X P_L = \sigma T_R^4 \left[(1-\alpha_W)A_W + (1-\alpha_C)A_C + (1-\alpha_L)A_L\right], \quad (2)$$

where $\eta_s$ is the scattered laser power fraction, $\eta_X$ is the laser-to-X-rays conversion efficiency, $P_L$ is the total laser power, $\sigma$ is the Stefan-Boltzmann constant, $T_R$, $A_{W,C,L}$ and $\alpha_{W,C,L}$ have been stated in Sec. 3. So the power fraction of the wall loss, capsule absorbed, and LEHs loss can be stated as a unified expression:

$$\eta_{W,C,L} = \frac{(1-\alpha_{W,C,L})A_{W,C,L}}{(1-\alpha_W)A_W + (1-\alpha_C)A_C + (1-\alpha_L)A_L}, \quad (3)$$

where the fraction absorbed by the capsule is just the conventional coupling efficiency from hohlraum to capsule defined in Ref.[48].

Another important issue limiting the hohlraum performance is the plasma filling from the ablation of high-$Z$ wall material[65]. When the plasma filling becomes serious, the laser absorption region shifts far from the wall and the hydrodynamic loss and the thin coronal radiative loss through LEHs increases rapidly. In particular, the hohlraum filling is believed to cause symmetry swings late in laser pulse that are detrimental to the symmetry control of the hot spot at a high convergence[32]. Usually, a filling model is utilized to evaluate the filling time $\tau_f$, which is defined the time it takes for the hot laser channel to be filled to an electron density $n_e$ approaching $0.1n_c$, where $n_c[\text{cm}^{-3}]=1.1\times10^{21}/\lambda^2[\mu\text{m}^2]$ is the critical density for a laser light of wavelength $\lambda$[66]. The plasma-filling model derived in Ref. [66] and [67] was just developed for simple on-axis laser-hohlraum geometry[65]. A uniform filling model has been extended to other hohlraums with arbitrary shape and filled gas in Ref.[42]:

$$\tau_f \propto A_{\text{loss}}^{0.29} \beta^{-1.33}, \qquad (4)$$

where $A_{\text{loss}} = (1-\alpha_W)A_W + (1-\alpha_C)A_C + (1-\alpha_L)A_L$ is defined as the equivalent power loss area, $\beta = A_W/V_f$, $A_W$ is the wall area as stated above, and $V_f$ is the residual hohlraum volume deducting the volume of laser channels and capsule from hohlraum volume.

The performance of diverse hohlraum configurations with different CCRs are tabulated in Table 1. The parameters adopted in simulation are exhibited in the table caption. It merits mentioning that the detailed layout of laser beams incidence into the NVCH is unclear, so the peak asymmetry in the NVCH is blank in the table. The data in Table 1 indicate that all the peak asymmetry values are no more than 0.2% inside the hohlraums with 6LEHs and single-cone laser beams, i.e., SH with CCR=5.1 (close to the golden ratio proposed in Ref. [41]), TACH with CCR=2.2 and TAEH with CCR=2.8. According to Eq. (3), a smaller wall area will lead to less wall loss and more fraction of capsule absorbed, namely coupling efficiency, under the conditions of the same LEHs and capsule. So the TAEH with smaller wall area could provide more drive on the capsule, simultaneously, keep a considerable wall volume to mitigate the LPIs issue.

TABLE 1. Performance of diverse hohlraum configurations, with the same capsule of 2.2 mm in diameter and 2.4 mm-diameter LEHs. The ratio of flux from the laser spots and the wall is set to 2. The time-averaged albedo of the wall and capsule are set to 0.8 and 0.3, respectively. The filling volume in Eq. (3) is assumed to be equal to the hohlraum volume. It is notable that the $\tau_{f\text{-Rev5}}$ in the table means the filling time of ignition target Rev5-CH (300 eV) of NIC. The filling time is normalized to $\tau_{f\text{-Rev5}}$. For the detailed layout of laser beams in the NVCH is unclear, the peak asymmetry inside the NVCH is blank.

| Performance | Criterion | NVCH (CCR=3.4) | SH (CCR=5.1) | TACH (CCR=2.2) | TAEH (CCR=2.8) |
|---|---|---|---|---|---|
| Peak asymmetry | RMS/% | — | 0.10 | 0.07 | 0.08 |
| | $\delta F$/% | — | 0.20 | 0.19 | 0.19 |
| Coupling efficiency | Wall area/cm$^2$ | 2.868 | 3.680 | 3.178 | 2.435 |
| | LEHs area/ cm$^2$ | 0.2139 | 0.2714 | | |
| | Wall loss/% | 65.52 | 66.08 | 62.72 | 56.31 |
| | LEHs loss/% | 24.43 | 24.37 | 26.78 | 31.38 |
| | Capsule absorbed/% | 10.05 | 9.55 | 10.50 | 12.31 |
| Plasma filling | Wall volume/ cm$^3$ | 0.3987 | 0.7378 | 0.3640 | 0.3534 |
| | Filling time/$\tau_{f\text{-Rev5}}$ | 1.40 | 2.44 | 1.13 | 1.47 |

Performance of assorted hohlraum configurations, i.e., the peak drive asymmetry, coupling efficiency, and filling time, are summarized in Fig.7. For comparison, the values of asymmetry and filling time are tuned to the same scale, multiplied by a factor of 10 and 2, respectively. The filling time is normalized to the one inside the ignition target Rev5-CH of NIC, parameters of which are from Ref.[9]. The simulations indicate that the peak drive symmetry inside the three-axis elliptical hohlraum (TAEH) with a case-to-capsule ratio (CCR) of 2.8 is comparable to the ones inside the spherical hohlraum (SH) with CCR=5.1 and three-axis cylindrical hohlraum (TACH) with CCR=2.2. The filling time of plasma affecting the LPIs is between those of SH and TACH, and about 1.5 times of that in the ignition hohlraum Rev5-CH of NIC and close to the one inside the near-vacuum cylindrical hohlraum (NVCH) with CCR=3.4. In particular, the coupling efficiency is about 22%, 29% and 17% higher than the one inside the NVCH, SH and TACH, respectively. This means that given total laser energy of 1.8 MJ[2,11] and the same energy absorbed by the capsule, about 330 kJ, 404 kJ and 265 kJ of energy will be saved utilizing the TAEH, respectively, when the scattered fraction of laser beams and the laser-to-X-rays conversion efficiency are the same.

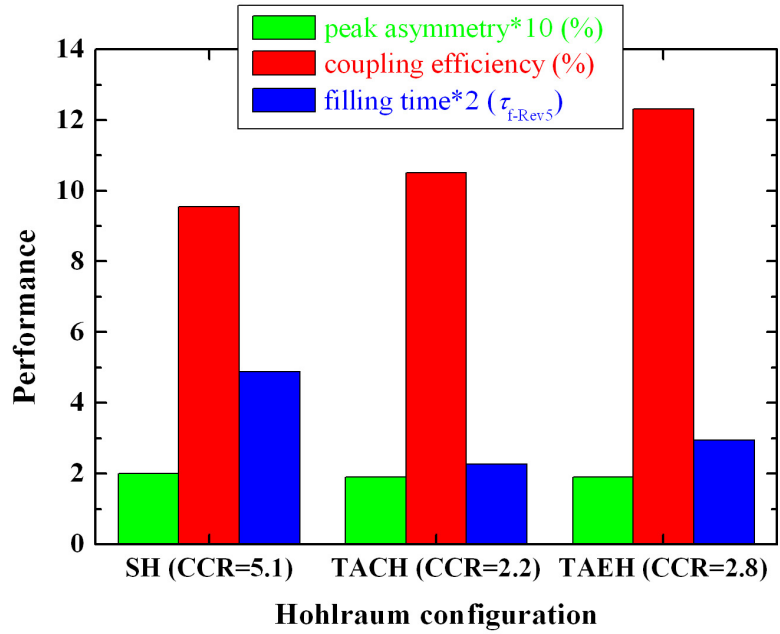

Figure 7. Tabulated performance of various hohlraum configurations. It should be notable that the values of performance criterions, except the coupling efficiency, are tuned to the same scale for comparison. The $\tau_{\text{f-Rev5}}$ in the legend means the filling time of ignition target Rev5-CH of NIC.

## 5. Discussions and conclusions

This proposed hohlraum here, together with the spherical hohlraum with six laser entrance holes and three-axis cylindrical hohlraum, have intrinsically better uniformity than traditional cylindrical hohlraum, but suffers the disadvantages of being fully three-dimensional, harder to model, and harder to manufacture. However, with advanced in modeling and manufacturing capabilities, these drawbacks may become minor. Besides, the hohlraum with three axis-elliptical cavities may be more convenient to be diagnosed experimentally, analogous to the case in the conventional cylindrical or elliptical hohlraum.

In summary, an advanced three-axis elliptical hohlraum (TAEH) is proposed to balance tradeoffs among the drive symmetry, coupling efficiency, and plasma filling of the hohlraum performance for indirectly driven inertial confinement fusion in this paper. Preliminary simulations from view-factor code IRAD3D indicate that the TAEH (with a case-to-capsule ratio, CCR=2.8) could provide excellent radiation symmetry during the thorough laser pulse of 'high-foot' drive, comparable to the ones inside the spherical hohlraum (SH) with CCR=5.1 and three-axis cylindrical hohlraum (TACH) with CCR=2.2. The filling time of plasma affecting the LPIs is between those of SH and TACH, and about 1.5 times of that in the ignition hohlraum Rev5-CH of NIC and close to the one inside the near-vacuum cylindrical hohlraum (NVCH) with CCR=3.4. In particular, the coupling efficiency is about 22%, 29% and 17% higher than the one inside the NVCH, SH and TACH, respectively. This means that given total laser energy of 1.8 MJ and the same energy absorbed by the capsule, about 330 kJ, 404 kJ and 265 kJ of energy will be saved utilizing the TAEH, respectively.

In conclusion, it would be envisioned that the proposed hohlraum configuration might be a viable approach worth pursuing for indirect-drive inertial confinement fusion, complementary to the conventional cylindrical hohlraum and the proposed recently novel hohlraums.

## Acknowledgements

This work was performed under the auspices of National Science Foundation of China under Grant Nos. 11475154, 51375185, U1430124, 11405160, 11505170 and 11304266.